\documentclass[aps,pre,preprint,groupedaddress]{revtex4-2}
\usepackage{amsmath}
\usepackage{graphicx}
\usepackage{bm}
\usepackage{hyperref}
\usepackage{physics}
\usepackage{lipsum} 

\newtheorem{identity}{Identity}
\newtheorem{simplification}{Simplification}
\newtheorem{definition}{Definition}[section]

\begin{document}

\title{Protein atom traps as seeing by neutron scattering}

\author{Georgii Koniukov}
\affiliation{University of Montpellier, 34090 Montpellier, France}

\begin{abstract}
The expression for the higher temperature dependence of the mean squared displacement in proteins is obtained. The quantum multi-well model explains the dynamic transitions of the proteins and minimizes the amount of parameters to a single one leading to the few-state harmonic system at low-temperatures, and, thus, justifying the possibility of using proteins as atom traps to control qubits at a bit higher temperatures.
\end{abstract}

\maketitle

\tableofcontents
\newpage

\section{Introduction}

In recent years, the study of protein dynamics has become an integral part of understanding biological processes at the molecular level. Proteins, being the workhorses of the cell, undergo a multitude of conformational changes that are crucial for their function. These dynamical behaviors of proteins are not merely incidental but are fundamental to their roles, from catalysis to structural support in the cellular matrix. Among various techniques employed to probe the dynamical properties of proteins, neutron scattering has emerged as a powerful tool, offering insights into the atomic and molecular motions within these complex macromolecules\cite{benedetto2011,gabel2002}.

Elastic Neutron Scattering (ENS) techniques, in particular, have been extensively utilized to investigate the Mean Square Displacement (MSD) of atoms in proteins, shedding light on their temperature-dependent dynamical behaviors. Benedetto (2011) showcased the application of ENS in studying the dynamical transition in lysozyme, highlighting the significant effect of hydration on the protein's dynamical properties and positing a relationship between the protein dynamical transition and the fragile-to-strong dynamical crossover (FSC)\cite{benedetto2011}. This study laid the groundwork for understanding the intricate relationship between protein dynamics, hydration, and temperature.

Building upon these foundational insights, Gabel et al. (2002) provided a comprehensive review of neutron scattering studies on protein dynamics, emphasizing the strong dependence of internal dynamics on the macromolecular environment. Their work underscored the versatility of neutron scattering in capturing the nuances of protein motions and their implications for biological function\cite{gabel2002}.

The intricate interplay between protein folding and dynamics was further elucidated by Nakagawa, Kamikubo, and Kataoka (2010), who investigated the effect of conformational states on the dynamical transition using incoherent elastic neutron scattering. Their findings revealed significant differences in the dynamical properties between wild-type proteins and mutants, suggesting that protein folding induces changes in dynamical behavior, a pivotal insight into the molecular basis of protein functionality\cite{nakagawa2010}.

Moreover, the studies by Nickels (2013) and Hirata (2018) explored the methodological aspects of neutron scattering in studying protein dynamics. Nickels (2013) focused on the impact of instrumental energy resolution on the analysis of MSD, while Hirata (2018) proposed a theoretical framework for interpreting the temperature dependence of MSD, shedding light on the physical underpinnings of the dynamical transition observed in proteins\cite{nickels2013,hirata2018}.

Collectively, these studies underscore the significance of neutron scattering techniques in unraveling the complex dynamics of proteins. By providing a detailed picture of how proteins move and respond to environmental changes, neutron scattering contributes to our understanding of the fundamental principles governing protein function. This body of work serves as a critical foundation for the current investigation, which aims to delve deeper into the dynamical transitions of proteins and their application.

In parallel, the advent of quantum computing has heralded a new era in the computational sciences, promising significant advancements in processing power and efficiency through the principles of quantum mechanics. Central to the development of quantum computers is the ability to control quantum states with high precision. Traditionally, this control has been achieved through the manipulation of atoms or ions in traps at ultracold temperatures, leveraging the quantum mechanical properties of these systems for computation. The literature is rich with studies demonstrating the efficacy of atom traps in quantum computing applications, highlighting advancements in laser cooling and trapping techniques, scaling of trapped ions, and optical trapping of Rydberg atoms among others\cite{ashkin1979,monroe2013,zhang2011}.

One of the foundational pieces of research by Ashkin and Gordon (1979) explored the potential of optical trapping and cooling of neutral atoms using resonance radiation pressure, setting the stage for subsequent developments in the field\cite{ashkin1979}. Miroshnychenko et al. (2006) further advanced this domain by demonstrating the precision manipulation of trapped atoms, a critical requirement for quantum information processing\cite{miroshnychenko2006}. The scalability of these systems, as discussed by Monroe and Kim (2013), remains a significant challenge, yet advancements in microfabricated ion traps and integrated photonics offer promising solutions\cite{monroe2013}.

Moreover, the application of blue-detuned optical traps proposed by Zhang, Robicheaux, and Saffman (2011) illustrates the versatility of trapping technologies in addressing the needs of quantum computing, particularly in achieving magic-wavelength conditions for various atomic states\cite{zhang2011}. Similarly, the innovative approach of using ferromagnetic nanowire domain walls as atom traps, as demonstrated by Allwood et al. (2006), underscores the potential for integrating quantum computing elements within semiconductor architectures\cite{allwood2006}.

However, a pervasive challenge in quantum computing remains the necessity of operating at ultracold temperatures to maintain coherence and minimize decoherence effects. This requirement imposes significant limitations on the practicality and scalability of quantum computing systems. It is within this context that the exploration of proteins as atom traps for controlling quantum states at higher temperatures emerges as a groundbreaking concept. Proteins, with their complex three-dimensional structures and functional versatility, offer a unique platform for quantum state manipulation. By leveraging the intrinsic properties of proteins, such as their ability to bind and interact with metal ions and small molecules, it is conceivable to create bio-inspired quantum computing architectures that operate at higher temperatures than currently possible with traditional atom traps.

This concept draws upon the extensive body of research on protein dynamics and their interactions with ligands and ions. For instance, the study of protein dynamics by neutron scattering techniques has provided valuable insights into the atomic and molecular motions within proteins, which could inform the design of protein-based quantum traps. Additionally, the manipulation and controlled assembly of atoms by optical tweezers, as explored by Miroshnychenko et al. (2006), could find analogous strategies in the biochemical manipulation of proteins and their bound atoms or molecules\cite{bustamante2020,ritchie2015}.

In proposing proteins as a novel platform for quantum computing, this paper seeks to bridge the gap between the physical sciences and biochemistry, opening new avenues for research in quantum information processing. By harnessing the natural properties of proteins and integrating them with existing quantum computing technologies, we aim to overcome the limitations imposed by ultracold temperatures and pave the way for more accessible and scalable quantum computing solutions. The potential for proteins to act as atom traps and control quantum states at higher temperatures not only represents a significant leap in the field but also underscores the interdisciplinary nature of future advancements in quantum computing.
\section{The Phenomenon}\label{phenomenon}

\begin{figure}[htb]
\includegraphics[width=5.6cm]{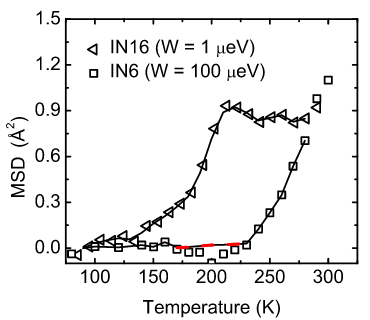}
\caption{Glutamate Dehydrogenase: MSD(T) dependence with possible saturation, \cite{vural}}
\label{fig:glu}
\end{figure}

\begin{figure}[htb]
\includegraphics[width=8.8cm]{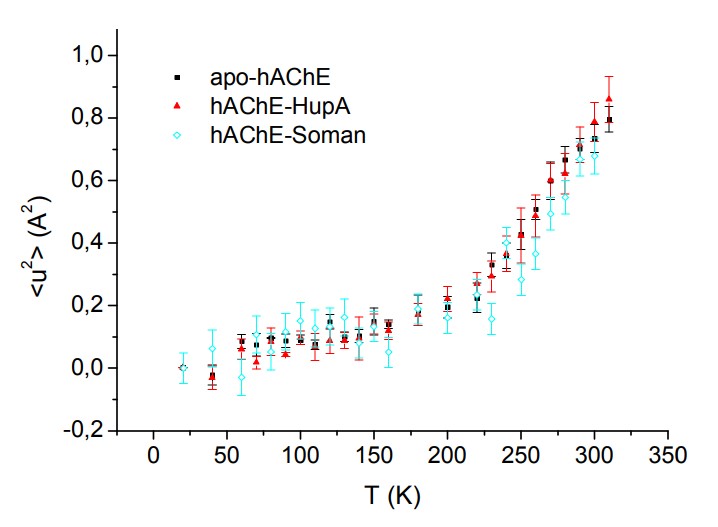}
\caption{Cholinesterase MSD(T) dependence with probable saturation, \cite{peters}}
\label{fig:cho}
\end{figure}

We aim to explain the behavior on Figures \ref{fig:GFP},\ref{fig:glu},\ref{fig:cho} (the plot of the MSD dependence of hydrogen atom w.r.t temperature) namely the zones:

\begin{figure}[htb]
\includegraphics[width=8.2cm]{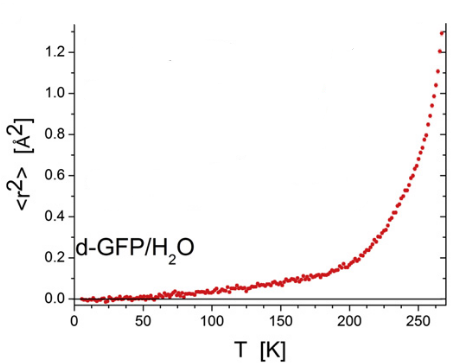}
\caption{Green Fluorescent Protein: MSD(T) dependence \cite{nickels}}
\label{fig:GFP}
\end{figure}
Regarding FIG. \ref{fig:GFP}:
\begin{enumerate}
  \item $\approx 0-50$K The temperature-independent region proposed by W. Doster \cite{doster}
  \item $\approx 50-190$K Almost linear dependence (similar to Brownian Motion)
  \item $\approx 190-280(?)$K Directed motion 
  \item Confined motion (Figure \ref{fig:cho}, when an resolution of the instrument is high enough to observe saturation)
\end{enumerate}

\section{The Probable Potential for the Hydrogen Atom}

\subsection{Two Types of Potential Wells}

Thousands of molecules surround the hydrogen atom in the hydrated proteins, thus, the effective potential well of such surroundings could have a too complicated shape to calculate MSD. Therefore as the first and rough approximation, we might look at such potential wells in a very general way. We might state that there are two general shapes:

\begin{enumerate}
  \item Square well, that could be used to tightly fill all the space as a sequence of square blocks.
  \item Non-square well.
\end{enumerate}

\begin{figure}[htb]
\includegraphics[width=4.3cm]{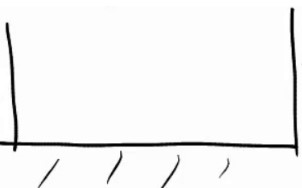}
\caption{Square Well}
\label{fig:lattice}
\end{figure}

\begin{figure}[htb]
\includegraphics[width=4.3cm]{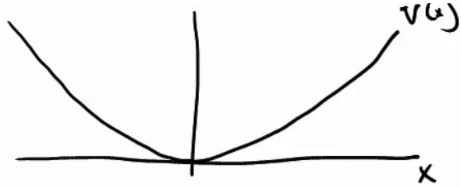}
\caption{Non-square Well}
\label{fig:nonlattice}
\end{figure}

\subsection{Approximation}

\subsubsection{Potential}

Then to start with, we may choose the simplest representatives of the two above:

\begin{enumerate}
  \item Infinite square well
  \item Harmonic oscillator
\end{enumerate}

\begin{figure}[htb]
\includegraphics[width=8.5cm]{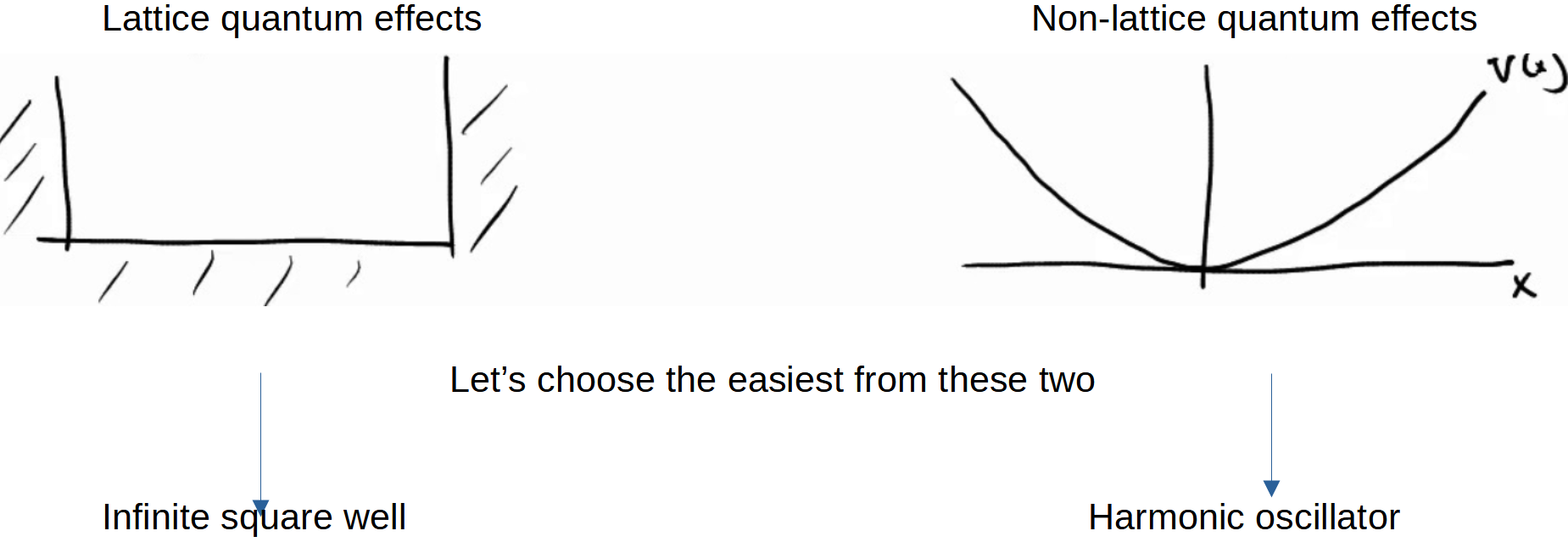}
\caption{Two types of potentials}
\label{fig:model}
\end{figure}

The harmonic oscillator already describes very well the zero-point fluctuations in the MSD of proteins, but
\begin{itemize}
  \item the harmonic oscillator does not describe the desired phase transition
  \item the harmonic oscillator is infinitely spread in space while the protein has a finite size, thus, one might try to limit the harmonic oscillator with the infinite square well, which also corresponds to the "narrow" harmonic oscillator:
\end{itemize}

\begin{figure}[htb]
\includegraphics[width=2.75cm]{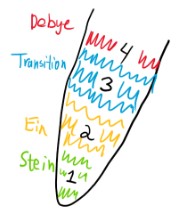}
\caption{The oscillator limited with infinite square well}
\label{fig:limitedOscillator}
\end{figure}

But this limitation will not lead to the phase transition of the protein on the Fig \ref{fig:cho}:

\begin{figure}[htb]
\includegraphics[width=5.0cm]{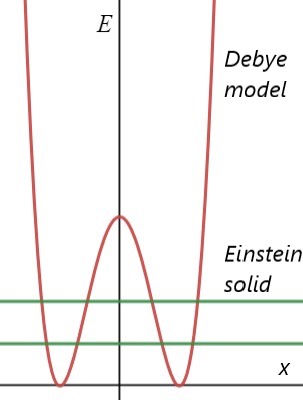}
\caption{Two oscillators limited with infinite square well}
\label{fig:twoLimitedOscillators}
\end{figure}

We will try to fit the results with two limited oscillators on the Fig \ref{fig:twoLimitedOscillators}. The reason for this amount is specified in Chapter \ref{chapter4} in Section \ref{answer}.

\subsubsection{Phase Transition}
The phase transition 2-3 in Chapter \ref{phenomenon} according to the model on Fig \ref{fig:twoLimitedOscillators} could be explained by the collective transitions of the hydrogens in the "harmonic" zone. Since in the Einstein model, all the oscillators are independent and have the same frequency $\omega$ they might change the energy level at the same condition when the temperature is

\begin{equation}\label{transitionCondition}
    kT^*=\hbar\omega
\end{equation}

After this jump to the higher energy level, the hydrogen atom might have more space to move (because of jumping above the barrier or tunneling) and, thus, a bigger mean square displacement in the phase transition 2-3. Thus, after this jump, the influence of the interaction between wells takes place.

The next jump at the temperature $kT^{**}=2\hbar\omega$ will increase the contribution of the infinite square well leading to the confined motion and saturation of the mean square displacement due to the finite size of the model.

We will first consider the two limiting cases separately: the infinite square well (high energies of the hydrogen) and the harmonic oscillator (low energies of the hydrogen) and then we will go to the direct calculation of the numerical values.

\section{Infinite Square Well}

\begin{figure}[htb]
\centering
\includegraphics[width=5cm,scale=1.0]{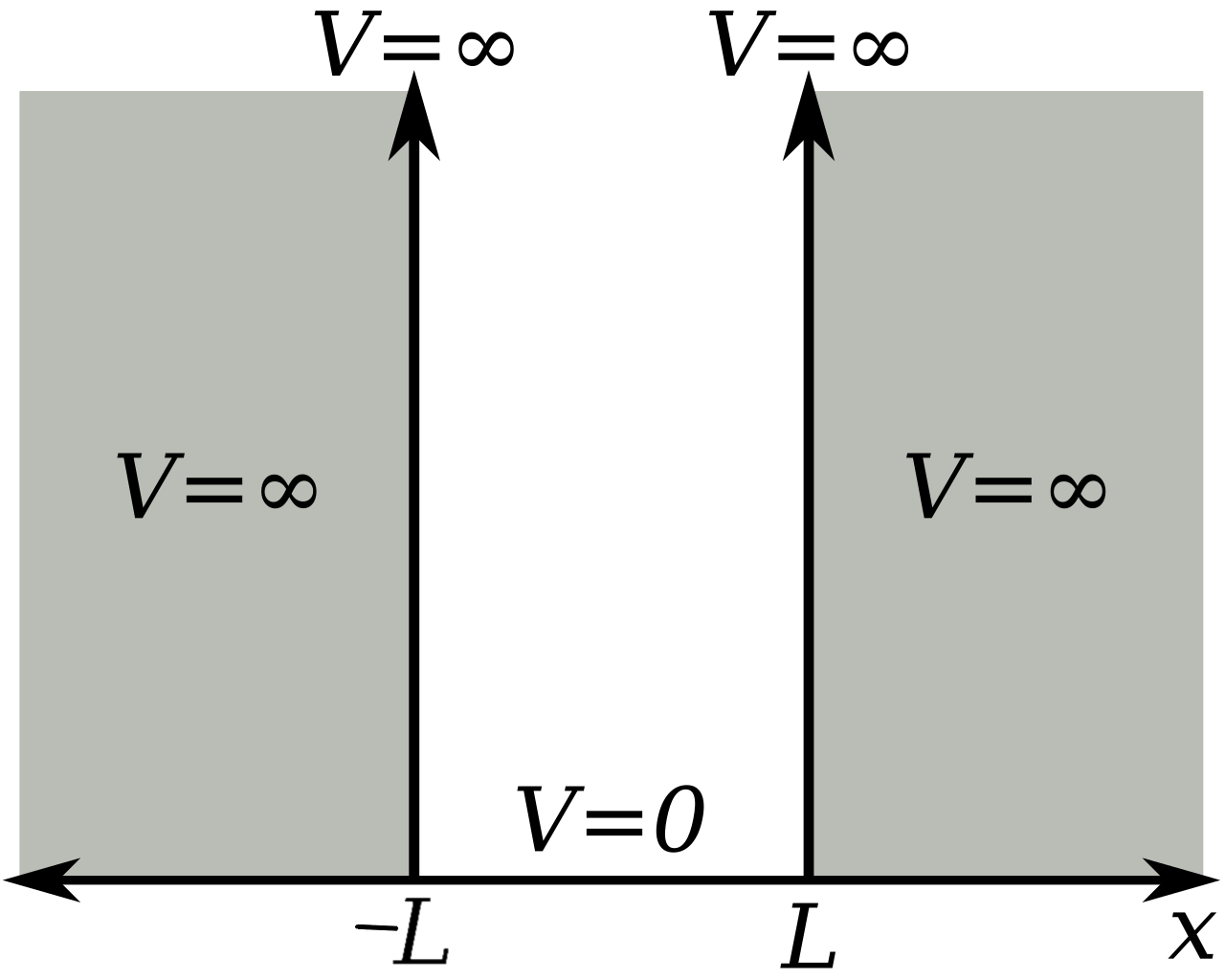}
    \caption{Infinite square well}
    \label{fig:infiniteWell}
\end{figure}
Let
\begin{equation}
    t^{**}\equiv \frac{\pi\hbar^2}{8kmL^2}\cdot\frac{1}{T^{**}}
\end{equation}
Where $k$ is the Boltzmann constant, $L$ is a half-width of the infinite square well and $m$ is a masss of the hydrogen.
\begin{widetext}
\begin{equation}
     \langle \Delta x^2 \rangle=\frac{L^2}{3}-\frac{2L^2}{\pi^2}\pi\cdot\frac{2\sqrt{t^{**}}-2\sqrt{t}+t-t^{**}+\frac{1}{\pi}\sum_{n=1}^\infty\frac{1}{n^2}\exp(-\pi t^{**} n^2)}{\frac{1}{\sqrt{t}}-1}
\end{equation}
\end{widetext}

\section{Harmonic Oscillator}

The Einstein solid is a model of a crystalline solid that contains a large number of independent three-dimensional quantum harmonic oscillators of the same frequency $\omega$.

\begin{equation}
    \langle \Delta x^2\rangle=\frac{\hbar}{2m\omega}\coth(\frac{\hbar\omega}{2kT})
\end{equation}
\section{Calculation of numerical values}\label{chapter4}
\subsection{Plan} 
\begin{enumerate}
    \item Knowing the temperature  $T^*$ of the phase transition 2-3 from the experimental data on the Fig \ref{fig:GFP}, we can define the frequency of the zero-point fluctuations $\omega$ ($\hbar\omega=kT^*$ according to the \ref{transitionCondition})$\iff$ knowing the fitting (value) the of $\langle \Delta x^2 \rangle\big|_{T=0}$  , we can estimate the frequency $\omega$ according to the \ref{msdOscillator} $\left(\frac{\hbar}{2m\omega}=\langle \Delta x^2\rangle\big|_{T=0}\right)$ and then the temperatures of the phase transitions $T^*$,  $T^{**}$ ($kT^{*}=\hbar\omega$, $kT^{**}=2\hbar\omega$).
    \item Then knowing the temperature $T^{**}$ and the mean square displacement at this temperature according to \ref{msdInfiniteSquareWell}, we can estimate the width of the square well $L$.
\end{enumerate}
\subsection{Parameters}
Thus, to do the calculation we have to know $T^*$ or $\langle \Delta x^2\rangle\big|_{T=0}$.
\subsection{Glutamate Dehydrogenase}
\begin{figure}[htb]
\includegraphics[width=5.6cm]{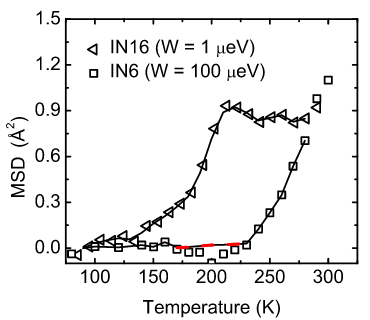}
\caption{Glutamate Dehydrogenase: MSD(T) dependence with possible saturation, \cite{vural}}
\label{fig:glu2}
\end{figure}

\subsubsection{Harmonic oscillator zone}
From the Fig. \ref{fig:glu2} (IN16 data) \cite{vural} we can notice the first saturation ordinate at: 
\begin{equation}
    T^{**}\approx210K;\quad \text{then} \ T^{*}=\frac{T^{**}}{2}=105K;
\end{equation}
And, thus:
\begin{equation}
    \omega = \frac{kT^*}{\hbar} \approx \frac{1.38\cdot10^{-23} J\cdot K^{-1}\cdot105 K}{1.054\cdot10^{-34}J\cdot s}\approx137.48\cdot10^{11}s^{-1}
\end{equation}
\begin{multline}
    \langle x^2\rangle\big|_{T=0}=\frac{\hbar}{2m\omega}=\frac{1.054\cdot10^{-34}J\cdot s}{2\cdot1.67\cdot10^{-27}kg\cdot137.48\cdot10^{11}s^{-1}}\\ \approx0.0023\cdot10^{-18}m^2=
    0.0023\ nm^2=0.23 \AA^2 
\end{multline}
That corresponds to the order of the experimental data
\subsubsection{Infinite square well zone}
Frome the \ref{twoStarsParameter}:
\begin{multline}
    t^{**}=\frac{\pi\hbar^2}{8km}\frac{1}{T^{**}}\frac{1}{L^2}=\\
    \frac{3.14\cdot(1.054\cdot10^{-34}J\cdot s)^2}{8\cdot1.38\cdot10^{-23} J\cdot K^{-1}\cdot1.67\cdot10^{-27}kg}\cdot\frac{1}{210K}\cdot\frac{1}{L^2}\\ \approx0.0009\cdot10^{-18} \ \frac{J\cdot s^2}{kg}\cdot\frac{1}{L^2}
\end{multline}
But $L^2>\langle x^2\rangle\big|_{T=T^{**}}$, because L is the infinite square well width, thus at least $L\geq 0.94 \AA$ (according to the data on the Fig. \ref{fig:glu2} and the fact that $L$ should be greater than the hydrogen atom radius):
\begin{multline}
    t^{**}=0.0009\cdot10^{-18} \ \frac{J\cdot s^2}{kg}\cdot\frac{1}{L^2}\leq \\ \leq
    0.0009\cdot10^{-18} \ \frac{J\cdot s^2}{kg}\cdot\frac{1}{0.9\cdot10^{-20} \ m^2}=0.1
\end{multline}
Thus, according to the \ref{tEstimation}:
\begin{equation}
        \begin{cases}
                \frac{1}{\pi}\sum_{n=1}^\infty\frac{1}{n^2}\exp(-\pi t^{**} n^2)<\frac{1}{\pi}\sum_{n=1}^\infty\frac{1}{n^2}=\frac{\pi}{6} \\
        \frac{1}{\pi}\sum_{n=1}^\infty\frac{1}{n^2}\exp(-\pi t^{**} n^2)>\frac{1}{\pi}\exp(-\pi t^{**}) \approx 0.23
    \end{cases}
\end{equation}
As a result the estimation:
\begin{equation}
    \frac{1}{\pi}\sum_{n=1}^\infty\frac{1}{n^2}\exp(-\pi t^{**} n^2)\approx \frac{1}{2}
\end{equation}
\begin{figure}[htb]
\includegraphics[width=8cm,scale=1.0]{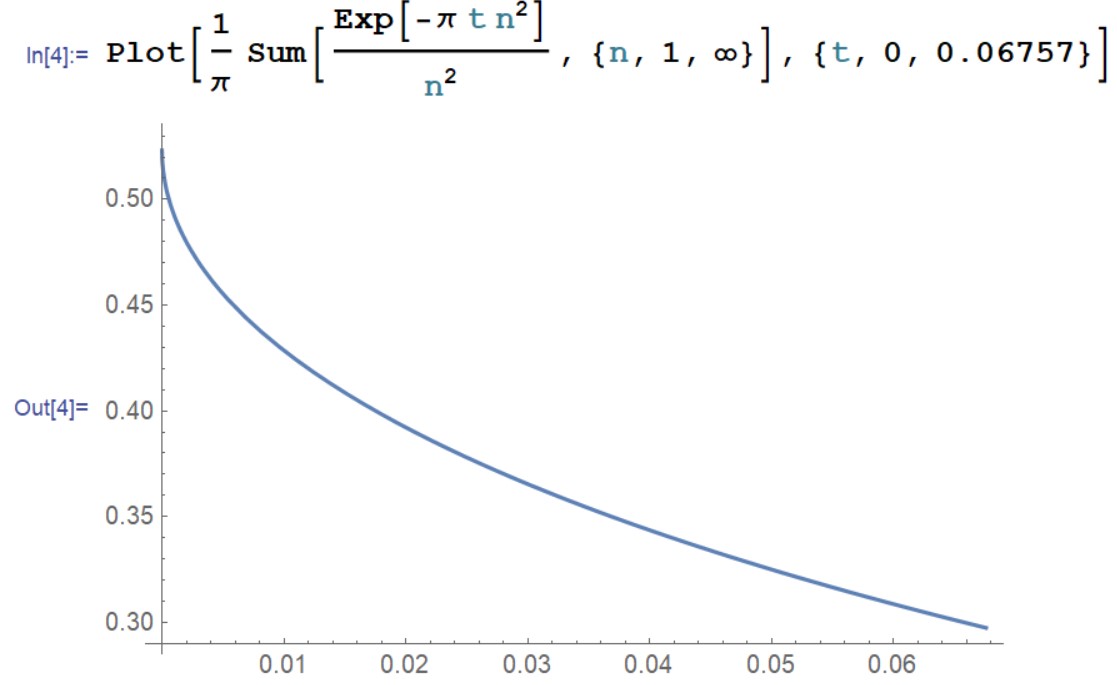}
    \caption{$\frac{1}{\pi}\sum_{n=1}^\infty\frac{1}{n^2}\exp(-\pi t^{**} n^2)$ w.r.t. $t^{**}$}
    \label{fig:tEstimation1}
\end{figure}
After substituting into the \ref{msdInfiniteSquareWell}:
\begin{widetext}
\begin{multline}
     \langle \Delta x^2\rangle = 
       \frac{L^2}{3}-\frac{2L^2}{\pi^2}\pi\cdot\frac{2\sqrt{t^{**}}-2\sqrt{t}+t-t^{**}+\frac{1}{\pi}\sum_{n=1}^\infty\frac{1}{n^2}\exp(-\pi t^{**} n^2)}{\frac{1}{\sqrt{t}}-1}=\\
      \frac{L^2}{3}-\frac{2L^2}{\pi^2}\pi\cdot\frac{2\sqrt{t^{**}}-2\sqrt{t}+t-t^{**}+\frac{1}{2}}{\frac{1}{\sqrt{t}}-1} 
\end{multline}
\end{widetext}
At $t=t^{**}$:
\begin{equation}
   \langle x^2 \rangle \big|_{T^{**}=210K}=0.9\AA^2=\frac{L^2}{3}-\frac{L^2}{\pi}\cdot\frac{1}{\frac{1}{\sqrt{t^{**}}}-1} \iff L\approx1.175\AA
\end{equation}
And the width of the infinite square well $2L=2.350\AA$.
\subsubsection{Harmonic Oscillator - Infinite Square Well contact}
\begin{figure}[htb]
\includegraphics[width=8.1cm,scale=1.0]{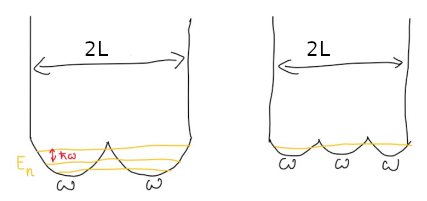}
    \caption{The more oscillators $-$ the less their independence and the less amount of harmonic levels contributes to the mean square displacement}
    \label{fig:harmonicOscillarorsInfiniteSquareWell}
\end{figure}
We will check how many energy levels of the harmonic oscillator are reachable before the infinite square well zone in case of two harmonic oscillators:
\begin{equation}
    W_p=\frac{m\omega^2x^2}{2}
\end{equation}
\begin{multline}
    W_p\big|_{x=-L}=W_p\big|_{x=L}=\frac{1.67\cdot10^{-27}kg\cdot(137.48\cdot10^{11}s^{-1})^2(1.175\AA)^2}{2} \\ =
    0.218\cdot10^{-20}J
\end{multline}
\begin{multline}\label{answer}
     n=\left[\frac{ W_p\big|_{x=-L}}{\hbar\omega}-\frac{1}{2}\right]=\left[\frac{ W_p\big|_{x=-L}}{\hbar\omega}-\frac{1}{2}\right]=\\\left[\frac{0.218\cdot10^{-20}J}{1.055\cdot10^{-34} J\cdot s \cdot 137.48\cdot10^{11}s^{-1} }-\frac{1}{2}\right]=1
\end{multline}
Thus, we have 2 energy levels in the harmonic zone (including zero-point energy). Therefore it is better to use two harmonic oscillator wells rather than a greater amount of wells to treat the independence effect in Einstein approximation.

\section{Conclusion}
The discovery of two harmonic zones within Glutamate Dehydrogenase presents an intriguing possibility for controlling the transitions between these zones, thereby allowing for the encoding of information. While the research to date has illuminated some aspects of these transitions, the exact temperature dependence at temperatures exceeding ${T^*}$ remains undetermined.

This research underscores the value of enhanced resolution in neutron scattering instruments, when instruments can accurately observe the saturation point of the mean squared displacement (MSD), and it is possible to ascertain the total number of harmonic zones accessible to the hydrogen atom within the protein structure.

Considering the temperature constraints inherent to qubit operation, due to entanglement phenomena at elevated temperatures, the harmonic wells of Glutamate Dehydrogenase at or near the temperature ${T^*}$ offer a promising avenue for qubit representation. The states of the hydrogen atom in these wells are particularly significant, as they contribute markedly to the protein's mean squared displacement under these conditions. This relationship between the protein's thermal state and the behavior of hydrogen atoms within its harmonic zones not only deepens our comprehension of protein dynamics but also opens up novel pathways for the utilization of protein states in quantum computing. The ability to leverage these harmonic zones for information storage and manipulation could herald a new era of bio-inspired quantum computing devices, where biological molecules and quantum computing mechanisms converge.
\section{Prospects}
One of the pivotal goals for forthcoming research could be the identification and analysis of other proteins exhibiting discernible saturation points in their mean square displacement (MSD) profiles when analyzed through neutron scattering techniques. 

Given the instrumental resolution effects on neutron scattering studies of protein dynamics as detailed by J.D. Nickels (2013) and D.Vural (2012) \cite{nickels2013,vural}, our approach will entail utilizing high-resolution neutron scattering instruments to enhance our ability to detect these critical saturation points. Such advancements will not only enhance our understanding of protein behavior under various thermal conditions but also facilitate the use of proteins as atom traps in quantum computing.

\appendix
\section{Appendix}
\subsection{Infinite Square Well}

The symmetric and asymmetric solutions:
\begin{equation}
\begin{cases}
    \braket{x}{n_s}=\psi_s(x)=\sqrt{\frac{1}{L}}\cos(\frac{\pi(2k+1)x}{2L})\\
    \braket{x}{n_a}=\psi_a(x)=\sqrt{\frac{1}{L}}\sin(\frac{\pi2kx}{2L})
\end{cases}
\end{equation}
\subsubsection{The MSD at T=0K}
The MSD for even and odd energy levels $n$:
\begin{widetext}
\begin{multline}
        \langle x^2 \rangle_{ns}\equiv\bra{n_s}x^2\ket{n_s}=\frac{1}{L}\int\limits_{-L}^Lx^2\left(\cos(\frac{\pi(2k+1)x}{2L})\right)^2\,dx=\frac{L^2}{3}-\frac{2L^2}{\pi^2n^2}
\\
        \langle x^2 \rangle_{na}\equiv\bra{n_a}x^2\ket{n_a}=\frac{1}{L}\int\limits_{-L}^Lx^2\left(\sin(\frac{\pi2kx}{2L})\right)^2\,dx =\frac{L^2}{3}-\frac{2L^2}{\pi^2n^2}
\end{multline}
\end{widetext}
Thus, the MSD for the symmetric and antisymmetric energy levels has the same expression.
\subsubsection{The MSD at $T\neq0$}
\begin{widetext}
    \begin{equation}\label{MSD:debye}
    \langle \Delta x^2\rangle = \frac{\sum\limits_{n=1}^\infty\bra{n}\Delta x^2\ket{n}e^{-\frac{E_n}{kT}}}{\sum\limits_{n=1}^\infty e^{-\frac{E_n}{kT}}}= \frac{\sum\limits_{n=1}^\infty\left(\frac{L^2}{3}-\frac{2L^2}{\pi^2n^2}\right)e^{-\frac{1}{kT}\frac{n^2\pi^2\hbar^2}{8mL^2}}}{\sum\limits_{n=1}^\infty e^{-\frac{1}{kT}\frac{n^2\pi^2\hbar^2}{8mL^2}} }
\end{equation}
\end{widetext}

We will try to simplify the expression \ref{MSD:debye} in order to find the temperature dependence of the $\Delta x^2$.
\\
\begin{definition}{Jacobi theta function}
\begin{equation}
    \Theta(t)=\sum\limits_{n=-\infty}^\infty\exp(-\pi n^2 t)
\end{equation}
\end{definition}
\begin{identity}{Jacobi's Identity}
\begin{equation}
    \Theta(t)=\frac{1}{\sqrt{t}}\Theta\left(\frac{1}{t}\right)
\end{equation}
For the relatively small values of $t\leq 0.01: \Theta(t)\approx \frac{1}{\sqrt{t}}$ \cite{lebedev}
\end{identity}
\begin{identity}
\begin{equation}
        \int\limits_t^\infty\exp(-\pi \tau n^2)d\tau=\frac{\exp(-\pi t n^2)}{\pi n^2}
\end{equation}
\end{identity}
\begin{simplification}{Partition function Z}
\begin{multline}
Z(T)=\sum\limits_{n=1}^\infty\exp(-\frac{1}{kT}\frac{n^2\pi^2\hbar^2}{8mL^2})\equiv \\ \sum\limits_{n=1}^\infty\exp(-\pi t n^2)=\frac{\Theta(t)-1}{2}
\end{multline}
Let
\begin{equation}\label{twoStarsParameter}
    t^{**}\equiv \frac{\pi\hbar^2}{8kmL^2}\cdot\frac{1}{T^{**}}
\end{equation}
Where $T^{**}$ is the temperature of the phase transition when the infinite square well approximation is starting to be applicable, because of the high population of the high energy levels.
\end{simplification}
\begin{simplification}
\begin{widetext}
\begin{multline}
        \sum\limits_{n=1}^\infty\frac{2L^2}{\pi^2n^2}\exp(-\frac{1}{kT}\frac{n^2\pi^2\hbar^2}{8mL^2})\equiv C\sum\limits_{n=1}^\infty\frac{1}{n^2}\exp(-\frac{1}{kT}\frac{n^2\pi^2\hbar^2}{8mL^2})\\ =C\sum\limits_{n=1}^\infty\frac{1}{n^2}\exp(-\pi t n^2)=C\pi\int\limits_t^\infty\sum\limits_{n=1}^\infty\exp(-\pi \tau n^2)d\tau \\ =C\pi\int\limits_t^\infty\sum\limits_{n=1}^\infty\exp(-\pi \tau n^2)d\tau=C\pi\int\limits_t^{t^{**}}\sum\limits_{n=1}^\infty\exp(-\pi \tau n^2)d\tau+C\pi\int\limits_{t^{**}}^{\infty}\sum\limits_{n=1}^\infty\exp(-\pi \tau n^2)d\tau\\
        =C\pi\int\limits_t^{t^{**}}\sum\limits_{n=1}^\infty\exp(-\pi \tau n^2)d\tau+const^{**}=C\pi\int\limits_t^{t^{**}}\frac{\Theta(\tau)-1}{2}d\tau+const^{**}
        \\ \approx C \pi \int\limits_t^{t^{**}}\frac{\frac{1}{\sqrt{\tau}}-1}{2}d\tau+const^{**}=C\frac{\pi}{2}\left(2\sqrt{t^{**}}-2\sqrt{t}+t-t^{**}\right)+const^{**}
\end{multline}
\end{widetext}
\end{simplification}
\begin{simplification}
\begin{widetext}
\begin{multline}\label{msdInfiniteSquareWell}
       \langle \Delta x^2\rangle = \frac{L^2}{3}-C\pi\cdot\frac{2\sqrt{t^{**}}-2\sqrt{t}+t-t^{**}+const^{**}}{\Theta(t)-1}\approx
       \frac{L^2}{3}-C\pi\cdot\frac{2\sqrt{t^{**}}-2\sqrt{t}+t-t^{**}+const^{**}}{\frac{1}{\sqrt{t}}-1}=\\
       \frac{L^2}{3}-\frac{2L^2}{\pi^2}\pi\cdot\frac{2\sqrt{t^{**}}-2\sqrt{t}+t-t^{**}+\int\limits_{t^{**}}^\infty\frac{(\Theta(\tau)-1)}{2}d\tau}{\frac{1}{\sqrt{t}}-1}
\end{multline}
\end{widetext}

\end{simplification}
\begin{simplification}
    \begin{multline}
        \int\limits_{t^{**}}^\infty\frac{\Theta(\tau)-1}{2}d\tau=\int\limits_{t^{**}}^\infty\sum_{n=1}^\infty\exp(-\pi\tau n^2)d\tau
        \\ =\frac{1}{\pi}\sum_{n=1}^\infty\frac{1}{n^2}\exp(-\pi t^{**} n^2)
    \end{multline}
    \begin{equation}\label{tEstimation}
    \begin{cases}
                \frac{1}{\pi}\sum_{n=1}^\infty\frac{1}{n^2}\exp(-\pi t^{**} n^2)<\frac{1}{\pi}\sum_{n=1}^\infty\frac{1}{n^2}=\frac{\pi}{6} \\
        \frac{1}{\pi}\sum_{n=1}^\infty\frac{1}{n^2}\exp(-\pi t^{**} n^2)>\frac{1}{\pi}\exp(-\pi t^{**})  
    \end{cases}
  \end{equation} 
\end{simplification}

\subsection{Harmonic Oscillator}
\subsubsection{The MSD}
\begin{equation}
    E_n=\hbar\omega\left(n+\frac{1}{2}\right)=\langle W_p\rangle +\langle W_k \rangle=2\langle W_p\rangle=m\omega^2\langle \Delta x^2\rangle_n
\end{equation}
\begin{equation}
    Z=\sum\limits_{n=0}^\infty\exp(-\frac{\hbar\omega\left(n+\frac{1}{2}\right)}{kT})=\frac{\exp(-\frac{\hbar\omega}{2kT})}{1-\exp(-\frac{\hbar\omega}{kT})}=\frac{1}{2\sinh(\frac{\hbar\omega}{2kT})}
\end{equation}
\begin{equation}
    \beta\equiv\frac{1}{kT}
\end{equation}
\begin{equation}
    \langle E\rangle=-\frac{1}{Z}\partial_\beta Z = \frac{\hbar\omega}{2}\coth(\frac{\hbar\omega}{2kT})
\end{equation}
\begin{equation}\label{msdOscillator}
    \langle \Delta x^2\rangle=\frac{\hbar}{2m\omega}\coth(\frac{\hbar\omega}{2kT})
\end{equation}


\begin{thebibliography}{99}
\bibitem{benedetto2011} Salvatore Magazu, Federica Migliardo, and Antonio Benedetto, \textit{Puzzle of Protein Dynamical Transition}, The Journal of Physical Chemistry B 115(24), (2011)
 
\bibitem{gabel2002} F. Gabel, D. Bicout, U. Lehnert, M. Tehei, M. Weik, G. Zaccai, \textit{Protein dynamics studied by neutron scattering}, Quarterly Reviews of Biophysics 35, (2002).

\bibitem{nakagawa2010} H. Nakagawa, H. Kamikubo, M. Kataoka, \textit{Effect of conformational states on protein dynamical transition}, Biochimica et Biophysica Acta (BBA) - Proteins and Proteomics 1804 1, (2010).

\bibitem{nickels2013} J. Nickels, \textit{Instrumental resolution effects in neutron scattering studies of protein dynamics}, Chemical Physics 424, (2013).

\bibitem{hirata2018} F. Hirata, \textit{On the interpretation of the temperature dependence of the mean square displacement (MSD) of protein, obtained from the incoherent neutron scattering}, Journal of Molecular Liquids, (2018).

\bibitem{ashkin1979} A. Ashkin, J. Gordon, \textit{Cooling and trapping of atoms by resonance radiation pressure}, Optics letters 4, (1979).

\bibitem{miroshnychenko2006} Y. Miroshnychenko, W. Alt, I. Dotsenko, L. Förster, M. Khudaverdyan, D. Meschede, D. Schrader, A. Rauschenbeutel, \textit{Quantum engineering: An atom-sorting machine}, Nature 442, (2006).

\bibitem{bustamante2020} C. Bustamante, L. M. Alexander, K. Maciuba, C. M. Kaiser, "Single-Molecule Studies of Protein Folding with Optical Tweezers," \textit{Annual Review of Biochemistry}, vol. 89, pp. 443-470, 2020.

\bibitem{ritchie2015} D. Ritchie, M. Woodside, "Probing the structural dynamics of proteins and nucleic acids with optical tweezers," \textit{Current Opinion in Structural Biology}, vol. 34, pp. 43-51, 2015.

\bibitem{monroe2013} C. Monroe, Jungsang Kim, \textit{Scaling the Ion Trap Quantum Processor}, Science 339, (2013).

\bibitem{zhang2011} S. Zhang, F. Robicheaux, M. Saffman, \textit{Magic-wavelength optical traps for Rydberg atoms}, Physical Review A 84, (2011).

\bibitem{allwood2006} D. Allwood, T. Schrefl, G. Hrkac, I. Hughes, C. Adams, \textit{Mobile atom traps using magnetic nanowires}, Applied Physics Letters 89, (2006).

















\bibitem{nickels} Jonathan D. Nickels, Hugh O’Neill, Liang Hong, Madhusudan Tyagi, Georg Ehlers, Kevin L. Weiss, Qiu Zhang, Zheng Yi, Eugene Mamontov, Jeremy C. Smith, Alexei P. Sokolov, \textit{Dynamics of Protein and its Hydration Water: Neutron Scattering Studies on Fully Deuterated GFP}, Biophysical journal (2012)
\bibitem{doster} W Doster, S Cusack, W Petry,
 \textit{Dynamical transition of myoglobin revealed by inelastic neutron scattering}, Nature (1989)
\bibitem{peters} Judith Peters,   Nicolas Martinez, Marie Trovaslet, Kévin Scannapieco, Michael Marek Koza, Patrick Massonde  and  Florian Nachoncd, \textit{Dynamics of human acetylcholinesterase bound to non-covalent and covalent inhibitors shedding light on changes to the water network structure}, Physical Chemistry Chemical Physics (2016).
\bibitem{vural} Derya Vural, Henry R. Glyde, \textit{Intrinsic mean-square displacements in proteins}, Phys. Rev. E 86, (2012)
\bibitem{lebedev} Lebedev N.N., \textit{Special functions and their applications} (1972)

\end{thebibliography}
\end{document}